\documentclass[twocolumn,showpacs,preprintnumbers,amsmath,amssymb]{revtex4}

\usepackage{graphicx}
\usepackage{dcolumn}
\usepackage{bm}
\usepackage{latexsym}

\begin{document}

\preprint{APS/paper}

\title{Phenomenological theory of phase transitions in epitaxial BaTiO$_3$ thin films}

\author{ V. B. Shirokov }
\author{ Yu. I. Yuzyuk }
\affiliation{Faculty of Physics, Southern Federal University, Zorge 5, Rostov-on-Don, 344090, Russia}
\author{ B. Dkhil }
\affiliation{Laboratoire Structures Propri\'et\'es et Mod\'elisation des Solides, Ecole Centrale Paris, UMR-CNRS 8580, F-92290 Chatenay-Malaby, France}
\author{ V. V. Lemanov }
\affiliation{A.F. Ioffe Physico-Technical Institute, 194021 St. Petersburg, Russia}

\date{\today}

\begin{abstract}
We develop a phenomenological thermodynamic theory of ferroelectric BaTiO$_3$ (BT) thin films epitaxially grown on cubic substrates using the Landau-Devonshire eight-order potential. The constructed "misfit-temperature" phase diagram is asymmetrical. We found that, overall view of the phase diagram depends on the values of compliances used in calculations and provide two qualitatively different diagrams. A thermodynamic path for BT film grown onto a particular substrate can be found using a plot of the room-temperature tetragonal distortion $ (c-a)/a $ as a function of misfit strain.
\end{abstract}

\pacs{64.60.Kw, 64.70.Kb, 77.65.Ly, 77.80.Bh }

\maketitle

\section{\label{sec:level1}INTRODUCTION}

Epitaxial ferroelectric BaTiO$_3$ (BT) films have been intensively investigated for their applications in high dielectric layers of capacitors, in dynamic random access memories (DRAM), microwave integrated circuits, and micro electro mechanical systems (MEMS) \cite{c1,c1a}. Nanoscale ferroelectric heterostructures lead to new phenomena. The physical properties of ferroelectric thin films substantially differ from those of bulk ferroelectrics. Epitaxially grown thin films are usually highly constrained because their fabrication is accompanied by several strain factors such as misfit strain due to lattice mismatch between the film and the substrate or the intermediate buffer layer, thermoelastic strain generated by the difference between the thermal-expansion coefficients (TECs) of the film and the substrate, self-strain of the paraelectric-ferroelectric phase transformation if the film is deposited above the Curie temperature ($T_c$), and defects such as dislocations and vacancies, appeared during the deposition. The effects of internal stresses on the electrical and electromechanical properties have been investigated experimentally and theoretically for a number of ferroelectric systems \cite{c2, c3, c4, c5, c6, c7, c8, c9, c10, c11, c12, c13, c14, c15, c16, c17, c18}. The experimental investigation of the above-mentioned factors is difficult because the physical properties of real thin films are a combined result of these factors. Therefore, theoretical analysis is important since it can provide fundamental insights into the behavior of thin films.

A phenomenological thermodynamic theory of ferroelectric BT thin films epitaxially grown on cubic substrates was developed \cite{c4} using the Landau-Devonshire six-order potential as a polynomial of the polarization components. For single-domain BaTiO$_3$ film, the "misfit-temperature" phase diagram was constructed. It was found that the two-dimensional (2D) clamping of the film, apart from a shift of the temperature of the ferroelectric transition, changes the entire phase diagram, creating new phases that are not present in bulk crystals. For BT thin films epitaxially grown on cubic substrates Pertsev \textit{et al} \cite{c4} have predicted five distinct stable phases. Four of them, namely $\bm{p}$, $\bm{c}$, $\bm{aa}$ and r have a common multiphase point on the diagram at zero misfit strain and temperature equal to the $T_c$ of the bulk material.

Later \cite{c5}, another phase diagram was proposed for BT film using another set of coefficients in the same six-order thermodynamic potential. In the vicinity of $T_c$ and small misfit strains, the phase diagram is quite similar to the previous one, except the low-temperature range where the 2D clamping stabilizes mainly the $\bm{r}$-phase and not the $\bm{ac}$ one. Recent first principle calculations \cite{c15,c17} qualitatively confirmed the latter finding and general view of the phase diagram. However, according to \cite{c17} the multiphase point has a positive shift along the misfit strain axis.

The misfit-temperature phase diagram constructed in \cite{c5}, is based on the six-order thermodynamic potential with the temperature-dependent coefficients \cite{c19,c20}, which reproduces quite well the low-temperature ferroelectric phase transitions in BT single crystal. It is worth noting that the 2D clamping increases considerably the ferroelectric phase transition temperature in perovskite thin films. In this case, the coefficient at the sixth-order term in thermodynamic potential becomes negative; therefore, the six-order expansion is not valid for high-temperature phase transitions in thin films. To overcome the problem, high-order terms should be included.
In this paper we employed an eighth-order polynomial for the Landau-Devonshire potential of bulk BT single crystal \cite{c21} under mechanical stress-free boundary conditions, only the first coefficient at $p^2$ is temperature dependent, while others are independent of temperature. We discuss temperature-misfit-strain phase diagrams for BT films on cubic substrates and their thermodynamic path determined using room-temperature tetragonal distortion value.

\section{A phenomenological thermodynamic potential}

Following Pertsev \textit{et al.} \cite{c4} we consider the thermodynamical phenomenological potential for BT thin film epitaxially grown in a cubic paraelectric phase on a thick cubic substrate. Standard Gibbs potential $\Phi (p,\sigma )$ is a function of polarization - $p$ and stress - $\sigma$ . Considering strain $u$ as an external parameter (either geometrical parameters or stresses are defined) the Helmholtz potential should be written as $F(p,u) = \Phi (p,\sigma ) + \sigma u$, where $\sigma = \sigma(u)$ can be found from the following requirement \cite{c22}:
\begin{equation}
\frac{\partial \Phi (p,\sigma )}{\partial \sigma } + u = 0.
\end{equation}

Starting from the Helmholtz potential $F(p,u)$ with mechanical forces or stresses as external parameters, the following expression is readily obtained for Gibbs potential $\Phi (p,\sigma ) = F(p,u) - \sigma u$ where $u=u(\sigma)$ can be found from the similar condition:

\begin{equation}
\frac{\partial F(p,u)}{\partial u} - \sigma  = 0.
\end{equation}

Equations (1) and (2) link the coefficients of the corresponding potentials. In the absence of external forces and at $p=0$, the strains $u$ are governed by the thermal expansion. As follows from Eq. (1), the Gibbs potential should contain the $-\alpha_0 (T-T_a)$  term, where  $\alpha_0$ is the TEC of the bulk material, $T_a$ is the temperature at which thermal deformations are absent. In the following we assume $T_a=0$.

The Helmholtz phenomenological eighth-order potential for cubic perovskite is given as \cite{c23,c21}:

\begin{widetext}
\begin{equation}
\begin{array}{l}
  F = \alpha _1^u (p_1^2  + p_2^2  + p_3^2 ) + \alpha _{11}^u (p_1^4  + p_2^4  + p_3^4 ) + \alpha _{12}^u (p_1^2 p_2^2  + p_1^2 p_3^2  + p_2^2 p_3^2 ) \qquad + G_6  \qquad + G_8  \hfill \\
\qquad - g_{11} (u_1 p_1^2  + u_2 p_2^2  + u_3 p_3^2 ) - g_{12} [u_1 (p_2^2  + p_3^2 ) + u_2 (p_1^2  + p_3^2 ) + u_3 (p_1^2  + p_2^2 )]  - g_{44} \left( {u_4 p_2 p_3  + u_5 p_1 p_3  + u_6 p_1 p_2 } \right) \hfill \\
\qquad + \frac{1}{2}c_{11} \left( {u_1^2  + u_2^2  + u_3^2 } \right)  + c_{12} \left( {u_1 u_2  + u_1 u_3  + u_2 u_3 } \right)

                  + \frac{1}{2}c_{44} \left( {u_4^2  + u_5^2  + u_6^2 } \right) - \alpha _0 T(c_{11}  + 2c_{12} )(u_1  + u_2  + u_3 ), \hfill \\
   \\
  G_6  = \alpha _{111} (p_1^6  + p_2^6  + p_3^6 ) + \alpha _{112} [p_1^4 (p_2^2  + p_3^2 ) + p_2^4 (p_1^2  + p_3^2 ) + p_3^4 (p_1^2  + p_2^2 )]  + \alpha _{123} p_1^2 p_2^2 p_3^2 , \hfill \\
   \\
  G_8  = \alpha _{1111} (p_1^8  + p_2^8  + p_3^8 ) + \alpha _{1112} [p_1^6 (p_2^2  + p_3^2 ) + p_2^6 (p_1^2  + p_3^2 ) + p_3^6 (p_1^2  + p_2^2 )]   + \alpha _{1122} (p_1^4 p_2^4  + p_1^4 p_3^4  + p_2^4 p_3^4 ) \hfill \\
\qquad + \alpha _{1123} (p_1^4 p_2^2 p_3^2  + p_1^2 p_2^4 p_3^2  + p_1^2 p_2^2 p_3^4 ) . \hfill \\
\end{array}
\end{equation}
\end{widetext}

The last term in the $F(p,u)$ expression (3) is responsible for the thermal expansion \cite{c22} of the cubic crystal. The strains $u_i$ are given in the Voigt matrix notation. The film is under short-circuit boundary conditions.

Here we consider (001) surface of a cubic substrate, which exhibits no symmetry change in the whole temperature range. The substrate defines the following strains at the film/substrate interface: $u_1 = u_2 = u_s, u_6 = 0$. They are governed by the symmetry of the substrate, the primary strain occurred at the film deposition, and TEC of the substrate. We assume the elastic fields to be homogeneous in the film volume and ignore any changes on the surface of the substrate. In the absence of external forces the remaining strains are determined by minimization of the Helmholtz potential (Eq. 3) with respect to the strains $u_i (i = 3, 4, 5)$,

\begin{equation}
\begin{array}{l}
  u_3  = \frac{g_{12} }{c_{11} }(p_1^2  + p_2^2 ) + \frac{g_{11} }{c_{11} }p_3^2  + \frac{c_{11}  + 2c_{12} }{c_{11} }\alpha _0 T - \frac{2c_{12} }{c_{11} }u_s  \hfill \\
  \quad = \left( Q_{12}  - \frac{s_{12} (Q_{11}  + Q_{12} )}{s_{11}  + s_{12} } \right)(p_1^2  + p_2^2 ) + \left( Q_{11}  - \frac{2s_{12} Q_{12} }{s_{11}  + s_{12} } \right)p_3^2    \hfill \\
  \quad + \frac{s_{11}  - s_{12} }{s_{11}  + s_{12} }\alpha _0 T - \frac{2s_{12} }{s_{11}  + s_{12} }u_s , \hfill \\
  u_4  = \frac{g_{44} }{c_{44} }p_2 p_3  = Q_{44} p_2 p_3 , \hfill \\
  u_5  = \frac{g_{44} }{c_{44} }p_1 p_3  = Q_{44} p_1 p_3 . \hfill \\
\end{array}
\end{equation}

Taking into account Eqs. 4 we can write the following thermodynamic potential of the thin film:

\begin{equation}
\begin{array}{l}
  G = \alpha _1 \left( {p_1^2  + p_2^2 } \right) + a_3 p_3^2 + \alpha _{11} \left( {p_1^4  + p_2^4 } \right) + a_{33} p_3^4  \hfill \\
  \qquad   + a_{12} p_1^2 p_2^2  + \alpha _{13} \left( {p_1^2 p_3^2  + p_2^2 p_3^2 } \right) \quad + G_6  \quad + G_8 . \hfill \\

\end{array}
\end{equation}

The following relations between coefficients of potential (5) and those in $F(p,u)$ and  $\Phi(p,\sigma )$ are (note, coefficients used in terms $G_6$ and $G_8$ are not changed under this transformation):

\begin{equation}
\begin{array}{l}
  \alpha _1  = \alpha _1^u  - \alpha _0 Tg_{12} \frac{{c_{11}  + 2c_{12} }}{{c_{11} }} - \left( {g_{11}  + g_{12} \frac{{c_{11}  - 2c_{12} }}{{c_{11} }}} \right)u_S \text{   } \hfill \\
  \qquad = \alpha _1^\sigma   - \frac{{Q_{11}  + Q_{12} }}{{s_{11}  + s_{12} }}\left( {u_S  - \alpha _0 T} \right), \hfill \\
  \alpha _3  = \alpha _1^u  - \alpha _0 Tg_{11} \frac{{c_{11}  + 2c_{12} }}{{c_{11} }} - 2\left( {g_{12}  - g_{11} \frac{{c_{12} }}{{c_{11} }}} \right)u_S \text{   } \hfill \\
  \qquad = \alpha _1^\sigma   - \frac{{2Q_{12} }}{{s_{11}  + s_{12} }}\left( {u_S  - \alpha _0 T} \right), \hfill \\
  \alpha _{11}  = \alpha _{11}^u  - \frac{{g_{12}^2 }}{{2c_{11} }} \hfill \\
  \qquad = \alpha _{11}^\sigma   + \frac{{(Q_{11}^2  + Q_{12}^2 )s_{11}  - 2Q_{11} Q_{12} s_{12} }}{{2(s_{11}^2  - s_{12}^2 )}}, \hfill \\
  \alpha _{33}  = \alpha _{11}^u  - \frac{{g_{11}^2 }}{{2c_{11} }} \hfill \\
  \qquad = \alpha _{11}^\sigma   + \frac{{Q_{12}^2 }}{{s_{11}  + s_{12} }}, \hfill \\
  \alpha _{12}  = \alpha _{12}^u  - \frac{{g_{12}^2 }}{{c_{11} }} \hfill \\
  \qquad = \alpha _{12}^\sigma   + \frac{{(Q_{11}^2  + Q_{12}^2 )s_{12}  - 2Q_{11} Q_{12} s_{11} }}{{(s_{11}^2  - s_{12}^2 )}} + \frac{{Q_{44}^2 }}{{2s_{44} }}, \hfill \\
  \alpha _{13}  = \alpha _{12}^u  - \frac{{g_{11} g_{12} }}{{c_{11} }} - \frac{{g_{44}^2 }}{{2c_{44} }} \hfill \\
  \qquad = \alpha _{12}^\sigma   + \frac{{Q_{12} (Q_{11}  + Q_{12} )}}{{s_{11}  + s_{12} }} + \frac{{Q_{44}^2 }}{{2s_{44} }} . \hfill \\
\end{array}
\end{equation}

Equations (6) evidently contains linear thermal expansion of the film, so that misfit strain  is $ u_m  = u_S  - \alpha _0 T $ \cite{c4}. Here $u_S$  contains the primary film strain $u_0$ and additional strain due to thermal shrinkage of the substrate appeared on cooling from the deposition temperature. The primary deformation appears due to misfit of the lattice parameters of the film and the substrate at the deposition temperature $T_0$. Starting from the deposition film can be under stress if $c$ (out-of plane) and $a$ (in-plane) parameters of the film at $T_0$ are different. Taking into account Eqs. (4) one can obtain the following expression:

\begin{equation}
u_0  \approx \frac{{c_{11} }}{{c_{11}  + 2c_{12} }}\frac{{a - c}}{c} = \frac{{s_{11}  + s_{12} }}{{s_{11}  - s_{12} }}\frac{{a - c}}{c},
\end{equation}
where $c$ and $a$ are the lattice parameters of the film at $T_0$.

Shrinkage of the substrate on cooling from $T_0$ down $T$ is obvious, $ \Delta  = b\left| {_T } \right. - b\left| {_{T_0 } } \right. = b_0 \alpha _S (T - T_0 ) $ , where  $\alpha_S$ is TEC of the substrate and $b_0$ the lattice parameter of the substrate at $T_0$. Then, film strain in the plane parallel to the substrate is

\begin{equation}
\begin{array}{l}
 u_S  = \frac{{a - a_0 }}{{a_0 }} = \frac{{a_0 (1 + \alpha _0 T_0  + u_0 ) + \Delta  - a_0 }}{{a_0 }}  \hfill \\
 \qquad = \alpha _0 T_0  + u_0  + \frac{{b_0 }}{{a_0 }}\alpha _S (T - T_0 ) .
\end{array}
\end{equation}

Therefore, $(u_s - \alpha_0 T)$ in Eqs. (6) is

\begin{equation}
u_m  = u_S  - \alpha _0 T = u_0  - (\alpha _0  - \frac{{b_0 }}
{{a_0 }}\alpha _S )(T - T_0 )
\end{equation}
In the following calculations we apply this to Eqs. (6).

\section{Phase diagram}

The phase diagram depends on the substrate used for film deposition because Eq. 9 contains lattice parameters and TEC of the substrate. In this case Eq. (9) determines the line that is the thermodynamic path for particular film/substrate heterostucture on the "temperature-misfit strain" ($T-u_m$) diagram. The slop of the thermodynamic path on the diagram is driven by the substrate/film lattice parameters ratio and by the difference of their TECs, so that
\[
u_m  = u_0  + (\alpha _0  - \frac{b_0 }{a_0 }\alpha _S )T_0.
\]

This approach is valid if and only if coefficients at $p^2$ are linear in $T$ and $u_m$. That is not true in the case of the sixth-order potential \cite{c19,c20}. One can expect rather small changes on the phase diagram when $(\alpha _0  - \frac{{b_0 }}{{a_0 }}\alpha _S )$ is small. However, we found the following quantitative changes of the $T - u_m$ phase diagram: (i) The stability range of the $\bm{ac}$ phase is limited in the low-temperature range when the difference between TEC of the film and the substrate is small; (ii) a tricritical point appears on the line between paraelectric and ferroelectric $\bm{c}$-phase.

\begin{figure}
\includegraphics{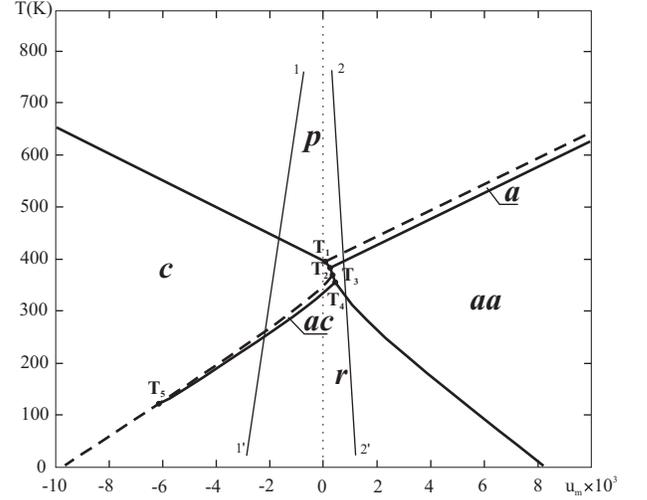}
\caption{\label{Fig.1} Temperature-misfit strain phase diagram BT film for the eight-order thermodynamic potential with the coefficients taken from column I in Table 1. Solid lines - first order phase transitions, dashed lines - second order phase transitions. The calculated coordinates of the multiphase points are $T_1: (u_m=0.01 \times 10^{-3}, T=388.2K)$, $T_2: (u_m=0.21 \times 10^{-3}, T=377K)$, $T_3: (u_m=0.31 \times 10^{-3}, T=366K)$, $T_4: (u_m=-0.39 \times 10^{-3}, T=354K)$, $T_5: (u_m=-6.20 \times 10^{-3}, T=114K)$. Thin lines $1-1^{'}$  and $2-2^{'}$  correspond to the thermodynamic path for BT film epitaxially grown with primary strain $u_0=0$ on MgO and SrTiO$_3$, respectively.}
\end{figure}

\begin{table}
\caption{\label{tab1}
Coefficients of the thermodynamical potential (5) for BaTiO$_3$, thin film calculated from the eighth-order potential (3) with the coefficients  taken from \cite{c21} $\alpha _1^\sigma   = 4.124 \times 10^5 (T - 402)$, and two sets of compliances taken from \cite{c24}-I (calculations) and \cite{c25}-II.
}
\begin{ruledtabular}
\begin{tabular}{cccc}
Constant& I\footnotemark[1] & II\footnotemark[2] & Units  \\ \hline \\
$\alpha_{11}$ & $2.56$ & $3.91$ & $\times 10^{8} Jm^5/C^4$ \\
$\alpha_{33}$ & $-0.298$ & $-0.05$ &  \\
$\alpha_{12}$ & $5.97$ & $4.13$ &  \\
$\alpha_{13}$ & $4.48$ & $4.00$ &  \\
$\alpha_{111}$ & $1.294$ & $1.294$ & $\times 10^{9} Jm^9/C^6$\\
$\alpha_{112}$ & $-1.95$ & $-1.95$ & \\
$\alpha_{123}$ & $-2.5$ & $-2.5$ & \\
$\alpha_{123}$ & $-2.5$ & $-2.5$ & \\
$\alpha_{1111}$ & $3.863$ & $3.863$ & $\times 10^{10} Jm^{13}/C^8$\\
$\alpha_{1112}$ & $2.529$ & $2.529$ & \\
$\alpha_{1122}$ & $1.637$ & $1.637$ & \\
$\alpha_{1123}$ & $1.367$ & $1.367$ & \\
$Q_{11}$ & $0.10$ & $0.10$ &  $m^4/C^2$\\
$Q_{12}$ & $-0.034$ & $-0.034$ &  \\
$Q_{44}$ & $0.029$ & $0.029$ &  \\
$s_{11}$ & $10.79$ & $8.33$ & $\times 10^{-12} m^2/N$\\
$s_{12}$ & $-4.365$ & $-2.68$ & \\
$s_{44}$ & $7.94$ & $9.24$ & \\
\end{tabular}
\end{ruledtabular}
\footnotetext[1]{$\alpha_1=4.124 \times 10^{5}(T-388-24.9 \times 10^3 u_m)$}
\footnotetext[1]{$\alpha_3=4.124 \times 10^{5}(T-388-28.3 \times 10^3 u_m)$}
\footnotetext[2]{$\alpha_1=4.124 \times 10^{5}(T-388+25.7 \times 10^3 u_m)$}
\footnotetext[2]{$\alpha_3=4.124 \times 10^{5}(T-388+29.2 \times 10^3 u_m)$ in $Jm/C^2$ units}
\end{table}

\begin{figure}
\includegraphics{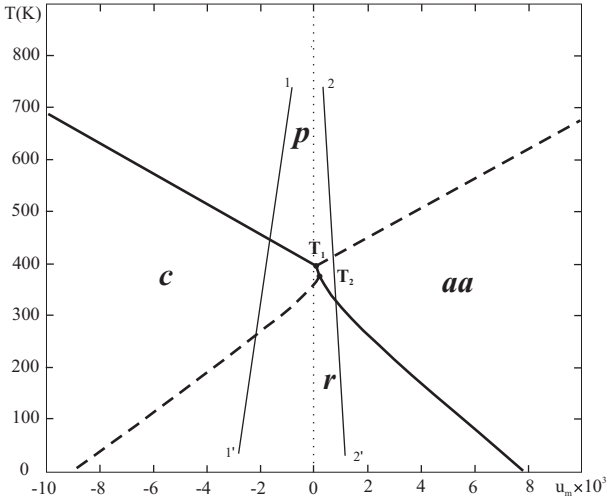}
\caption{\label{Fig.2}
Temperature-misfit strain phase diagram BT film for the eight-order thermodynamic potential with the coefficients taken from column II in Table 1. Solid lines - first order phase transitions, dashed lines - second order phase transitions. The calculated coordinates of the multiphase points are $T_1: (u_m=0.0002 \times 10^{-3}, T=388K)$, $T_2: (u_m=0.20 \times 10^{-3}, T=373K)$. Thin lines $1-1^{'}$  and $2-2^{'}$  correspond to the thermodynamic path for BT film epitaxially grown with primary strain $u_0=0$ on MgO and SrTiO$_3$, respectively.
}
\end{figure}

It is worth noting that values of the coefficients in  Eq. (5), and therefore, overall view of the phase diagram depends on the values of elastic moduli or compliances used in the calculations. Elastic moduli  in potential (3) should be taken at D = const, therefore, we used $c_{i,j}^D $\cite{c24} measured at room temperature. The corresponding calculated compliances are given in the second column of the Table 1. On the other hand, available compliances measured \cite{c25}, in the high-temperature paraelectric phase are slightly different. We used coefficients for eighth-order potential from Ref. \cite{c21}. According to expression (6) two sets of elastic moduli produce two sets of coefficients  of the thermodynamical potential for thin film listed in Table 1. Below we provide phase diagrams for two sets of coefficients.

The temperature-misfit strain phase diagram of BT film for the eighth-order potential with the coefficients taken from column I in Table 1 is shown in Fig. 1. Among eight allowed by symmetry phases, only six are present on the diagram,

$\bm{p} - \bm{p}=(0, 0, 0) P4/mmm (N123)$ paraelectric,

$\bm{c} - \bm{p}= (0, 0, p) P4mm (N99)$, polarization normal to the film surface,

$\bm{a} - \bm{p}= (p, 0, 0) Pmm2 (N25)$, polarization along the axis in plane of the film,

$\bm{aa} - \bm{p}= (p, p, 0) Amm2 (N38)$, polarization along the diagonal in plane of the film,

$\bm{r} - \bm{p}= (p_1, p_1, p_2) Cm (N8)$, polarization with $\bm{aa}$ and $\bm{c}$ components
\footnote{%
It is worth nothing that the $\bm{r}$ - phase is monoclinic one}

$\bm{ac} - \bm{p}= (p_1, 0, p_2) Pm (N6)$, polarization with $\bm{a}$ and $\bm{c}$ components.

The phase diagram for the eighth-order potential (Table 1, column I) contains one more phase - $\bm{a}$ with the polarization along one of the former cubic axis in plane of the film. This phase does not exist in the sixth-order Pertsev's model \cite{c5}. As shown in Fig.1, the $\bm{a}$-phase is stable in the narrow range between paraelectric and orthorhombic $\bm{aa}$ phase. Note, $\bm{p} - \bm{a}$ transition is second order, while $\bm{a} - \bm{aa}$ is first order. Another important difference with respect to \cite{c5} is the first order nature of the $\bm{p} - \bm{c}$ transition with very small hysteresis. Also, in contrast to Pertsev's model \cite{c5} the second order $\bm{c} - \bm{r}$ transition is possible only at very low temperature below $T_5$ point $(u_m= -6.20 \times 10^{-3}, T=114K)$. Above $T_5$ transition to the $\bm{r}$-phase is possible through the intermediate $\bm{ac}$-phase as a result of the second order $\bm{c} \to \bm{ac}$ transition, and following the first order $\bm{ac} \to \bm{r}$ one. As one can see in Fig. 1, the boundary between $\bm{c}$ and $\bm{a}$ phase in between $T_1$ and $T_2$ points, between $\bm{c}$ and $\bm{aa}$ in between $T_2$ and $T_3$, between $\bm{ac}$ and $\bm{aa}$ in between $T_3$ and $T_4$ corresponds to the first order transitions. Note, transition between $\bm{r}$ and $\bm{aa}$ is always the first order. Finally, the $T_1$ point $(u_m=0.01 \times 10^{-3}, T=388.2K)$ is slightly shifted with respect to zero misfit strain because in contrast to Pertsev's model \cite{c5} $\bm{p} - \bm{c}$ transition is of first order one. Such a positive shift along the misfit strain axis is in a qualitative agreement with the first principle calculations \cite{c17}.

\begin{figure}
\includegraphics{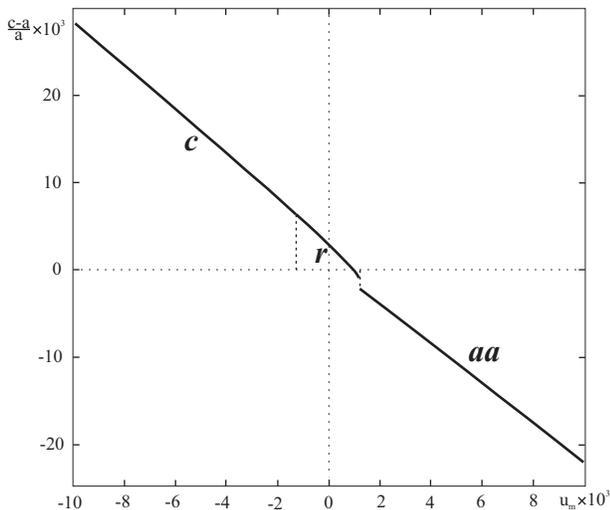}
\caption{\label{Fig.3}
Room-temperature tetragonal distortion $(c-a)/a$ as a function of misfit strain for eight-order thermodynamic potential with the coefficients taken from column I in Table 1.
}
\end{figure}

The temperature-misfit strain phase diagram of the BT film for the eighth-order potential with the coefficients taken from column II in Table 1 is shown in Fig. 2. Two intermediate phases $\bm{a}$ and $\bm{ac}$ disappear and only four phases $\bm{p}$, $\bm{c}$, $\bm{aa}$ and $\bm{r}$ are present on the diagram. The $\bm{p}$ - $\bm{c}$ transition is of first order-type, but hysteresis is smaller. Therefore, shift of the $T_1$ point with respect to zero deformation is now less pronounced.

Thin lines $1-1^{'}$  and $2-2^{'}$  in Figs. 1 and 2 correspond to the thermodynamic path ( Eq.(9) ) for BT film epitaxially grown at $T_0 = 1023K$  with primary strain $u_0=0$ on MgO and SrTiO$_3$, respectively. The primary strain $u_0$ depends on the growing conditions and formation of misfit dislocations in the film volume. Real thermodynamic path is shifted with respect to $1-1^{'}$  and $2-2^{'}$  lines on the actual $u_0$ value, which is determined from experiment.

Actual $u_0$ value can be found from the structural data. Figure 3 shows room-temperature tetragonal distortion $(c-a)/a$ as a function of misfit strain $u_m$ calculated using the following expression:

\begin{equation}
\begin{array}{l}
\frac{{c - a}}{a} = \left( {Q_{12}  - \frac{{s_{12} (Q_{11}  + Q_{12} )}}{{s_{11}  + s_{12} }}} \right)\left( {p_1^2  + p_2^2 } \right) \\
\qquad + \left( {Q_{11}  - \frac{{2s_{12} Q_{12} }}{{s_{11}  + s_{12} }}} \right)p_3^2  - \frac{{s_{11}  - s_{12} }}{{s_{11}  + s_{12} }}u_m.
\end{array}
\end{equation}

Using the experimental $(c-a)/a$ ratio and the plot in Fig.3, one can find room-temperature $u_m$ value that determines the corresponding point on the phase diagram (Figs. 1 or 2). The distance between this point and $1-1^{'}$ $(2-2^{'})$ line at $T=300K$ is equal to $u_0$.

\section{Conclusions}
We develop phase diagrams for epitaxial BT films on cubic substrates as a function of the misfit strain based on the eight-order Landau-Devonshire phenomenological potential, which is valid for high-temperature states. In the frame of this model the phase transition from the paraelectric to the ferroelectric $\bm{c}$-phase is always of first-order-type. This result has important consequence, the phase diagram is no more symmetrical with respect to the zero misfit strain axis.

Available from the literature, two sets of elastic moduli for bulk BT yields different phase diagrams with a different number of stable phases. Therefore, phase transition sequence in epitaxial thin films is governed not only by the misfit strain value $u_m$. Other important factors are thermodynamic conditions which can change chemical composition of the film. In the case of ferroelectric perovskites, oxygen pressure during film deposition seems to be a very important factor, because even weak deviation of the film density markedly changes its elastic properties.

The performed calculations allow to estimate possible phase transition temperatures for BT films on different substrates. We propose a simple way how to find a thermodynamic path on the phase diagram for BT film grown on MgO or SrTiO$_3$ substrate using room-temperature structural data.

\begin{acknowledgments}
This study was partially supported by the Russian Foundation for Basic Research Contract No. 06-02-16271
\end{acknowledgments}


\end{document}